\documentclass[a4paper,aps,12pt,nofootinbib,showpacs]{revtex4}
\usepackage{amsmath}
\usepackage{amssymb}
\usepackage{amsfonts}
\usepackage{cases}

\begin{document}
%%%%%%%%%%%%%%%%%%%%%%%%%%%%%%%%%%%%%%%%
\renewcommand{\include}[1]{}
\renewcommand\documentclass[2][]{}
\newcommand{\del}[1]{\partial_{#1}}
\newcommand{\stp}[2]{#1 \ \stackrel{{\rm S}}\otimes \ #2}
\newcommand{\K}[1]{\stackrel{(#1)}K}
\newcommand{\T}[2]{\stackrel{(#2)}#1}
\newcommand{\pbr}[2]{\bigl\{ \, #1 \ ,\ #2 \, \bigr\} _{{\rm PB}}}
\newcommand{\sbr}[2]{\bigl[ \, #1 \ ,\ #2 \, \bigr] _{{\rm S}}}
\newcommand{\kphi}{(K \, \partial \, \Phi)}
\newcommand{\xiphi}{(\xi \, \partial \, \Phi)}
\newcommand{\ip}[2]{#1 \, \cdot \, #2 }
%%%%%%%%%%%%%%%%%%%%%%%%%%%%%%%%%%%%%%%%
\newcommand{\eqname}{Killing hierarchy}

%%%%%%%%%%%%%%%%%%%%%%%%%%%%%%%%%%%%%%%%
\title{
Constants of Motion for Constrained Hamiltonian Systems \\
{\it --- A Particle around a Charged Rotating Black Hole ---}\\
}
\today
\hfill{OCU-PHYS 330}

\hfill{AP-GR 76}


\author{${}^{1}$Takahisa Igata} 
\email{igata@sci.osaka-cu.ac.jp}
\author{${}^{2}$Tatsuhiko Koike}
\email{koike@phys.keio.ac.jp}
\author{${}^{1}$Hideki Ishihara}
\email{ishihara@sci.osaka-cu.ac.jp}

\affiliation{%
${}^{1}$Department of Mathematics and Physics,
Graduate School of Science, Osaka City University,
Osaka 558-8585, Japan
\\
${}^{2}$Department of Physics, Keio University, Yokohama 223-8522, Japan
}

%%%%%%%%%%%%%%%%%%%%%%%%%%%%%%%%%%%%%%%%
\begin{abstract}
\vspace{1cm}
\centerline{\bf Abstract}
We discuss constants of motion of a particle under an external field in a curved
spacetime, taking into account the Hamiltonian constraint which arises
from reparametrization invariance of the particle orbit. As the necessary and
sufficient condition for the existence of a constant of motion, we obtain a set
of equations with a hierarchical structure, which is understood as a generalization
of the Killing tensor equation. 
It is also a generalization of the conventional argument in
that it includes the case when the conservation condition holds only on
the constraint surface in the phase space. In that case, it is shown that 
the constant of motion is associated with a conformal Killing tensor.  
We apply the hierarchical equations and find constants of motion
in the case of a charged particle in an electro-magnetic field in black hole
spacetimes. We 
also demonstrate that gravitational and electro-magnetic fields exist in which a
charged particle has a constant of motion associated with a conformal Killing
tensor.  
\end{abstract}
%%%%%%%%%%%%%%%%%%%%%%%%%%%%%%%%%%%%%%%%
\maketitle

%%%%%%%%%%%%%%%%%%%%%%%%%%%%%%%%%%%%%%%%%%%%%%%%%%%
\section{Introduction}
%%%%%%%%%%%%%%%%%%%%%%%%%%%%%%%%%%%%%%%%%%%%%%%%%%%

Black holes have great importance in modern physics. 
In astrophysics, black holes are thought to be central engines of active galactic nuclei. The gravitational and electro-magnetic fields of black holes play crucial roles for production of the energy actually observed.
On the other hand, higher-dimensional black holes, in recent years, 
gather much attention in the context of unified theories of interactions. 
An important task at present is to reveal their properties 
because existence of the extra dimensions could be verified by observations of 
the phenomena concerned with black holes.

A test particle is an important probe of black hole spacetimes 
because the study of motions of a test particle 
gives an important insight into physical properties of black holes. 
In the case of a charged test particle, 
the motion gives information on both 
the gravitational and electro-magnetic fields. 
The constants of motion, {\it i.e.}, the 
conserved quantities along a particle trajectory, 
are useful for analysis of the motion of the particle.
One can conclude that the equations of motion are integrable 
if one knows sufficient numbers of constants of motion satisfying a 
certain property. 

In the case of a particle without charge when the trajectory is a geodesic on a
curved spacetime,  
existence of a one-parameter group of 
isometries generated by a Killing vector implies the existence 
of a constant of motion which is linear in the momentum. 
The constants of motion which is non-linear in the momentum arise from Killing 
tensors.
For example, in the Kerr spacetime, 
there exists a constant of motion quadratic in the momentum \cite{Carter:1968rr} 
arises from a Killing tensor of rank 2 \cite{Walker:1970un}.
Furthermore, the existence of a rank-2 Killing tensor 
was recently shown in the Kerr-NUT-de Sitter black holes in any dimensionality 
\cite{Frolov:2007nt,Kubiznak:2006kt,Page:2006ka,Frolov:2006pe,
Krtous:2006qy,Krtous:2007xf}.

A constant of motion which is quadratic in the momentum for a charged particle 
in the Kerr-Newman spacetime was also found \cite{Carter:1968rr} 
by the method of Hamilton-Jacobi equation. 
It was shown that the constant of motion is related to a rank-2 Killing tensor 
also in this case \cite{Hughston:1972qf, P.Sommers:1973qf}, 
and a set of coupled equations is obtained which should be satisfied by 
the constants of motion for the charged particle \cite{P.Sommers:1973qf} 
(see also \cite{vanHolten:2006xq} for recent works).

The motion of a test particle in a curved spacetime is 
described by a world line with arbitrary parametrization. 
This reparametrization invariance gives rise to the Hamiltonian constraint. 
In this paper, 
we discuss a generalization of the conservation condition 
for the system of a particle which is subject to external fields 
in the Hamiltonian formalism. 
We consider the conservation condition with the constraint taken into account. 
Namely, we require that the conservation equation hold 
under the constraint condition. 
As a result, we obtain a generalized set of equations 
which is the necessary and sufficient condition for 
the existence of a constant of motion for a particle in an external field. 
The equations have hierarchical structure, and the topmost equation in the
hierarchy is the conformal Killing tensor equation. 

As applications, we first consider systems of a charged particle in
electro-magnetic 
fields around black holes, namely, a test Maxwell field 
on the Kerr spacetime, the Kerr-Newman spacetime, 
and a five-dimensional charged rotating black hole. 
In these cases, the Hamiltonian 
constraint does not play any role in finding constants of motion 
since the metrics admit rank-2 Killing tensors. 
In the final example, we demonstrate that 
a constant of motion can exist which is related to a conservation condition
holding only on the constraint surface. 
The example is constructed 
by conformal transformation 
of a spacetime with a test Maxwell field.

The organization of the paper is as follows. 
In the following section, we review the relation between constants of motion of a 
free particle and geometrical quantities by using the Hamiltonian formalism. 
In section \ref{sec:sec3}, we formulate the condition for existence of 
a constant of motion for a particle in an external field when there is 
a constraint condition. 
We obtain a set of equations with a hierarchical structure as a result. 
The equations are applied to systems of a charged particle in section
\ref{sec:sec4}. 
Finally, section \ref{sec:sec5} is devoted to a summary.

%%%%%%%%%%%%%%%%%%%%%%%%%%%%%%%%%%%%%%%%%%%%%%%%%%%%%%%%%%%%%%%%%%%%%%%%%%%%%%
\section{Conserved quantities of a free particle in the Hamiltonian Formalism}
\label{sec2}
%%%%%%%%%%%%%%%%%%%%%%%%%%%%%%%%%%%%%%%%%%%%%%%%%%%%%%%%%%%%%%%%%%%%%%%%%%%%%%

In this section, 
we review the relation between geometrical quantities of a curved spacetime 
$({\cal M}, g_{\mu \nu})$ and conserved quantities of a free particle 
in $({\cal M}, g_{\mu \nu})$. 
It is well known that the solutions of Killing equation, Killing fields, give 
conserved quantities along the trajectory of the particle, 
which is a geodesic. 
We shall derive the relation by using 
the Hamiltonian formalism. 
The relation will be generalized in the following section. 

Let $H$ be the Hamiltonian of a free particle given by 
\begin{eqnarray}
        H= \frac{1}{2 m}\left( g^{\mu\nu} p_\mu p_\nu + m^2 \right), 
\label{eq:geodesicHamiltonian}
\end{eqnarray}
where $m$ is the mass and $p_{\mu}$ is the canonical momentum of the particle. 
The Hamilton equation for \eqref{eq:geodesicHamiltonian} leads to 
the geodesic equation. 
Let $F$ be a dynamical quantity of the free particle represented by a function 
on the phase space with coordinates $(x^{\mu}, p_\nu)$. 
In the Hamiltonian formalism, 
if $F$ is a constant of motion, {\it i.e.}, a conserved quantity along the orbit of the particle, 
it commutes with $H$ under the Poisson bracket, 
\begin{eqnarray}
        \frac{dF}{d\tau} = \pbr{F}{H}
        := \frac{\partial F}{\partial x^\mu} \frac{\partial H}{\partial p_\mu} 
        -\frac{\partial H}{\partial x^\mu} \frac{\partial F}{\partial p_\mu} 
        =0,
\label{eq:conservation}
\end{eqnarray}
where the bracket with ${\rm PB}$ denotes the Poisson bracket, 
and $\tau$ is the proper time of the particle. 

Here, we assume that $F$ is written in the form
\begin{equation}
        F(x^\mu, p_\mu)= \xi^\mu p_\mu, 
\label{eq:xip}
\end{equation}
where $\xi^{\mu}$ is a vector field 
on ${\cal M}$. 
For $F$ in the form of \eqref{eq:xip}, the equation \eqref{eq:conservation} 
takes the form 
\begin{eqnarray}
        \pbr{F}{H}
        = \frac{1}{m} \xi^{\mu; \nu} p_{\mu} p_{\nu} = 0,
\end{eqnarray}
where the semicolon denotes the covariant derivative. 
Hence we have the {\it Killing equation} 
\begin{eqnarray}
        \xi^{\mu;\nu} = 0. 
\end{eqnarray}
A solution $\xi^\mu$ of the equation, a {\it Killing vector}, 
gives a constant of motion $F$.

We can generalize the Killing equation to a higher-rank tensor equation. 
Let us assume that $F$ has the form of
\begin{eqnarray}
        F = \K{k}{}^{\mu_1 \cdots \mu_k} p_{\mu_1}\cdots p_{\mu_k}, 
\end{eqnarray} 
where $\K{k}$ denotes a completely symmetric tensor field 
of rank $k$. 
Then the equation \eqref{eq:conservation} yields the Killing tensor equation 
for rank-$k$ tensor
\begin{eqnarray}
        \K{k}{}^{(\mu_{1}\cdots \mu_{k};\mu_{k+1})} = 0. 
\end{eqnarray}
That is, Killing tensors,
which are geometrical quantities, 
are related with constants of motion.

%%%%%%%%%%%%%%%%%%%%%%%%%%%%%%%%%%%%%%%%%%%%%%%%%%%%%%%%%%%%%%%%%%%%
\section{Formulation of Generalized Killing Equations}
\label{sec:sec3}
%%%%%%%%%%%%%%%%%%%%%%%%%%%%%%%%%%%%%%%%%%%%%%%%%%%%%%%%%%%%%%%%%%%%

In this section, we generalize the Killing tensor equation 
derived in the previous section 
to include the case when the conservation condition holds 
only on the constraint surface in the phase space.

Let us first derive the Hamiltonian for a free particle. 
The geodesic is obtained from the variational principle 
with the action being the arc length of a curve connecting two points, 
\begin{equation}
        S= -m \int_{\cal C} \sqrt{g_{\mu\nu} \frac{d x^\mu}{d\lambda}
        \frac{d x^\nu}{d\lambda}} ~d\lambda. 
\label{eq:geodesic}
\end{equation}
The action $S$ 
has the reparametrization invariance $\lambda \to \lambda'$. 
Introducing a Lagrange multiplier $N(\lambda)$, 
we can construct an equivalent action in the quadratic form, 
\begin{equation}
        S= \int_{\cal C} \left(\frac{m}{2} g_{\mu\nu}
\frac{d x^\mu}{Nd\lambda}\frac{d x^\nu}{Nd\lambda} - \frac{m}{2}\right) N
d\lambda. 
\label{eq:equiv}
\end{equation}
By the standard procedure of moving from the Lagrange formalism to 
the Hamilton formalism, we can derive the Hamiltonian, 
\begin{eqnarray}
        H= \frac{N}{2m}\left( g^{\mu\nu} p_\mu p_\nu + m^2\right). 
\label{eq:Hgeo}
\end{eqnarray}
Variation by $N$ yields the mass shell condition
or the Hamiltonian constraint, 
\begin{equation}
        {\cal H}(x^\mu, p_\mu)= g^{\mu\nu} p_\mu p_\nu + m^2 \approx 0. 
\label{eq:Hamiltonian_Constraint}
\end{equation}
The last equality, {\it weak equality}, defines a constraint surface 
in the phase space where the actual particle motion is confined 
on the hypersurface.
The action \eqref{eq:geodesicHamiltonian} is restricted to be used with 
an affine parameter, while an arbitrary parameter is allowed in \eqref{eq:equiv}.

Let us generalize the Killing tensor equation, taking into account 
the case when the conservation condition holds only on the constraint surface. 
We begin with a Hamiltonian slightly more general than \eqref{eq:Hgeo}, 
\begin{eqnarray}
        H=\frac{N}{2m} \left( g^ {\mu \nu}p_{\mu} p_{\nu} 
                + B^{\rho}p_{\rho}+V \right),
\label{eq:generalH}
\end{eqnarray}
to describe a system of a particle in an external field, 
where $B^{\rho}$ and $V$ are a vector field 
and a scalar field,  
respectively, on ${\cal M}$. 
By setting the differentiation of $H$ by $N$ equal to zero, we get 
the Hamiltonian constraint equation
\begin{eqnarray}
        {\cal H}= g^{\mu \nu}p_\mu p_\nu 
                + B^\rho p_\rho+V \approx 0. 
\label{eq:generalconstraint}
\end{eqnarray}

We shall examine the conservation condition for a function $F(x^\mu, p_\mu)$ 
on the constrained system. 
Since the particle motion is realized only on the constraint surface, 
then 
it suffices that $F$ commutes with $H$ only on the constraint surface, {\it
  i.e.}, 
\begin{eqnarray}
        \pbr{F}{H} = N\pbr{F}{{\cal H}} \approx 0.
\label{eq:constraintsurface}
\end{eqnarray}
This is equivalent to 
\begin{eqnarray}
        \pbr{F}{{\cal H}} + \phi {\cal H} = 0,
\label{eq:weakconservation}
\end{eqnarray}
where $\phi$ is an arbitrary function on the phase space.

We assume that $F$ and $\phi$ are expanded
in the form, 
\begin{align}
        F &=\sum_k \K{k}{}^{\mu_1 \cdots \mu_k}
                p_{\mu_1}\cdots p_{\mu_k} 
        =: \sum_k \ip{{\K{k}}}{p^k}, 
\label{eq:F}
\\
        \phi &=\sum_l \T{\lambda}{l}{}^{\mu_1 \cdots \mu_l}
                p_{\mu_1}\cdots p_{\mu_l} 
        =: \sum_l \ip{\T{\lambda}{l}{}}{p^l}, 
\label{eq:phi}
\end{align}
where 
$\K{k}{}^{\mu_1 \cdots \mu_k}$ 
and $\T{\lambda}{l}{}^{\mu_1 \cdots \mu_l}$ 
are symmetric tensor fields 
of rank $k$ and rank $l$, 
respectively, on ${\cal M}$. 
The right-hand sides of \eqref{eq:F} and \eqref{eq:phi} are 
abbreviations of contraction with $p$'s. 
Substituting these expressions into \eqref{eq:weakconservation}, we have 
\begin{align}
        &\sum_{k}\biggl(
                -\sbr{\K{k-1}}{\T{g}{2}}
                -\sbr{\K{k}}{\T{B}{1}}
                -\sbr{\K{k+1}}{\T{V}{0}} 
\cr     
        &\hspace{2cm}   +\stp{\T{\lambda}{k-2}}{\T{g}{2}}
                +\stp{\T{\lambda}{k-1}}{\T{B}{1}}
                +\T{\lambda}{k} \T{V}{0}
        \biggr) \cdot p^{k} =0,
\label{eq:pregeneralHEs}
\end{align} 
where we understand that $\K{k} =0$ and $\T{\lambda}{k} = 0$ for $k < 0$, 
and $\stackrel{{\rm S}}\otimes$ denotes symmetric tensor product. 
The bracket with the subscript ${\rm S}$ is the Schouten bracket \cite{Schouten}, 
which is the map from symmetric tensor fields 
$\stackrel{(k)}X$ and 
$\stackrel{(l)}Y$ of rank $k$ and rank $l$, respectively, 
to a rank-$(k+l-1)$ symmetric tensor field 
$\sbr{\T{X}{k}}{\T{Y}{l}}$ defined by 
\begin{align}
        \ip{ \sbr{\T{X}{k}}{\T{Y}{l}} }{p^{k+l-1}}
                =- \pbr{\ip{\T{X}{k}}{p^k}}{\ip{\T{Y}{l}}{p^l}}. 
\end{align}
Since \eqref{eq:pregeneralHEs} should be satisfied for 
any $p^{k}= p_{\mu_1}p_{\mu_2} \cdots p_{\mu_k}$, 
then the each coefficient of $p^{k}$ vanishes, {\it i.e.}, 
\begin{align}
        &\sbr{\K{k-1}}{\T{g}{2}}
        +\sbr{\K{k}}{\T{B}{1}}
        +\sbr{\K{k+1}}{\T{V}{0}}
\cr     
        &\hspace{2cm}   -\stp{\T{\lambda}{k-2}}{\T{g}{2}}
        -\stp{\T{\lambda}{k-1}}{\T{B}{1}}
        -\T{\lambda}{k} \T{V}{0} =0. 
\label{eq:general_hierarchy}
\end{align}
Let us call the set of equations \eqref{eq:general_hierarchy} 
the {\it \eqname} because it is 
a generalization of the Killing equation. 

When the highest rank of the hierarchy is $N$, namely, 
when $\K{l+1}=0$ and $\T{\lambda}{l}=0$ for $l \geq N$, the {\eqname} reads 
\begin{align}
        &- \sbr{\K{N}}{\T{g}{2}} 
        +\stp{\T{\lambda}{N-1}}{\T{g}{2}}
        = 0,
\label{eq:highestHE}
\\
        &- \sbr{\K{N-1}}{\T{g}{2}} 
        -\sbr{\K{N}}{\T{B}{1}}
        +\stp{\T{\lambda}{N-2}}{\T{g}{2}} 
        +\stp{\T{\lambda}{N-1}}{\T{B}{1}}
        = 0, 
\\
        &- \sbr{\K{k-1}}{\T{g}{2}} 
        -\sbr{\K{k}}{\T{B}{1}} 
        -\sbr{\K{k+1}}{\T{V}{0}}
\cr     
        &\hspace{2cm}   +\stp{\T{\lambda}{k-2}}{\T{g}{2}} 
        +\stp{\T{\lambda}{k-1}}{\T{B}{1}} 
        +\T{\lambda}{k} \T{V}{0} 
        = 0, \quad 2\leq k \leq N-1, 
\\
        &- \sbr{\K{0}}{\T{g}{2}} 
        -\sbr{\K{1}}{\T{B}{1}} 
        -\sbr{\K{2}}{\T{V}{0}}
        +\T{\lambda}{0} \T{B}{1} 
        +\T{\lambda}{1} \T{V}{0} 
        = 0, 
\\
        &- \sbr{\K{0}}{\T{B}{1}} 
        -\sbr{\K{1}}{\T{V}{0}} 
        +\T{\lambda}{0} \T{V}{0} 
        = 0, 
\label{eq:mainHE}
\end{align}
where $\ip{\T{g}{2}}{p^2}$ is assumed to be nonvanishing. 
The structure of the {\eqname} tells us that we should solve them 
from the highest-rank equation \eqref{eq:highestHE}. 
Since the highest-rank equation is the conformal Killing equation, 
existence of a conformal Killing tensor is necessary for 
a constant of motion to exist. 
If the {\eqname} admits a non-trivial solution, 
then there exist 
a constant of motion associated with a conformal Killing tensor. 

In the case of a free particle, {\it i.e.}, 
$\T{B}{1}=0$ and $\T{V}{0}=const.$,  
the {\eqname} reduces to a set of decoupled conformal Killing equations 
if the particle is massless, 
$\T{V}{0}=0$, and to a set of decoupled Killing equations if the particle is
massive, $\T{V}{0}=m^2$. This fact is shown in appendix \ref{sec:app1}. 

In the case of $\T{\lambda}{k}=0$ for all $k$ 
the {\eqname} reduces to
\begin{align}
        &- \sbr{\K{N}}{\T{g}{2}} = 0,
\cr
        &- \sbr{\K{N-1}}{\T{g}{2}} 
        -\sbr{\K{N}}{\T{B}{1}}
        = 0, 
\cr&
        - \sbr{\K{k-1}}{\T{g}{2}} 
        -\sbr{\K{k}}{\T{B}{1}} 
        -\sbr{\K{k+1}}{\T{V}{0}}
        = 0, \quad 2\leq k \leq N-1,
\cr&
        - \sbr{\K{0}}{\T{g}{2}} 
        -\sbr{\K{1}}{\T{B}{1}} 
        -\sbr{\K{2}}{\T{V}{0}}
        = 0, 
\cr&
        - \sbr{\K{0}}{\T{B}{1}} 
        -\sbr{\K{1}}{\T{V}{0}} 
        = 0.
\label{eq:zerolambdaHE}
\end{align}
These equations were obtained by Sommers \cite{P.Sommers:1973qf} 
and van Holten \cite{vanHolten:2006xq}. 
Several applications are found in \cite{Hierarchy}.

%%%%%%%%%%%%%%%%%%%%%%%%%%%%%%%%%%%%%%%%%%%%%%%%%%%%%%%%
\section{{\eqname} for a charged particle}
\label{sec:sec4}
%%%%%%%%%%%%%%%%%%%%%%%%%%%%%%%%%%%%%%%%%%%%%%%%%%%%%%%%

In this section, we apply the {\eqname} to the systems 
of an electrically charged particle subject to an external electro-magnetic
field as an important application of our formalism. 
We consider the Hamiltonian of a charged particle in the form
\begin{align}
        H = \frac{N}{2m}
\left[
        g^{\mu \nu}(p_{\mu}-q A_{\mu} )(p_{\nu}-q A_{\nu}) + m^2
\right],
\label{eq-charged-ham}
\end{align}
where $m$ and $q$ are the mass and the electric charge of the particle, 
respectively, 
and $A_{\mu}$ denotes the gauge potential. 
Substituting 
\begin{align}
        B^{\mu} = - 2 q A^{\mu}, \quad
        V=q^2 A_{\mu}A^{\mu} + m^2
\end{align}
in \eqref{eq:highestHE}-\eqref{eq:mainHE}, 
we obtain the {\eqname}
for a charged particle: 
\begin{align}
&
        - \sbr{\K{N}}{g} +\stp{\T{\lambda}{N-1}}{g}= 0, 
\label{eq:EMHEtop}
\\&
        - \sbr{\K{N-1}}{g} +2q\sbr{\K{N}}{A}
        +\stp{\T{\lambda}{N-2}}{g} 
        -2q\stp{\T{\lambda}{N-1}}{A} = 0, 
\\&
        - \sbr{\K{k-1}}{g} 
        +2q\sbr{\K{k}}{A} 
        -q^2\sbr{\K{k+1}}{A^2}
\\
&\hspace{1cm}
        +\stp{\T{\lambda}{k-2}}{g} 
        -2q\stp{\T{\lambda}{k-1}}{A} 
        +\T{\lambda}{k} (q^2 A^2 + m^2) 
        = 0, 
\quad 2\leq k \leq N-1,
\\&
        - \sbr{\K{0}}{g} 
        +2q\sbr{\K{1}}{A} 
        -q^2 \sbr{\K{2}}{A^2}
        -2q\T{\lambda}{0} A 
        +\T{\lambda}{1} (q^2 A^2 + m^2) 
        = 0, 
\\&
        2q \sbr{\K{0}}{A} 
        -q^2 \sbr{\K{1}}{A^2} 
        +\T{\lambda}{0} (q^2 A^2 + m^2) 
        = 0, 
\label{eq:EMHE}
\end{align}
where $A^2$ denotes the squared norm of $A^{\mu}$. 
Summing up all equations contracted by $A$'s, we have
\begin{eqnarray}
        m^2 ( q^{N-1} \ip{\T{\lambda}{N-1}}{A^{N-1}} 
                + q^{N-2}\ip{\T{\lambda}{N-2}}{A^{N-2}} 
                + \cdots + q \ip{\T{\lambda}{1}}{A^1} + \T{\lambda}{0})=0. 
\label{eq:generalconsistencycondition}
\end{eqnarray}
One can also derive \eqref{eq:generalconsistencycondition} 
by setting $p_\mu=qA_\mu$ in 
\eqref{eq:weakconservation} 
after calculating the Poisson bracket.

We note that existence of constants of motion linear in momenta 
for a massive particle requires 
existence of a Killing vector. 
This is so because, when $N=1$ and $m\neq 0$, 
the equation 
\eqref{eq:generalconsistencycondition} 
reduce to $\T{\lambda}{0} = 0$,  so that the highest-rank equation 
\eqref{eq:EMHEtop} 
becomes 
the Killing vector equation.

In what follows, we first 
consider three spacetimes with an electro-magnetic field: 
test electro-magnetic fields called the Wald solutions 
on the Kerr background, 
the four-dimensional Kerr-Newman black hole, 
and the five-dimensional charged rotating black hole. 
Since these systems admit a rank-2 Killing tensor $\K2$, 
we consider rank-2 solutions of the {\eqname} 
with $\T{\lambda}{l}=0$, 
\begin{align}
&
- \sbr{\K{2}}{g} =0,
\label{eq:N=2EMHEtop}
\\&
        - \sbr{\K{1}}{g} + 2q \sbr{\K{2}}{A} =0,
\label{eq:N=2EMHEmiddle}
\\&
        - \sbr{\K{0}}{g} + 2 q \sbr{\K{1}}{A} 
        -q^2 \sbr{\K{2}}{A^2} =0.
\label{eq:N=2EMHEbottom}
\end{align}
Next, in the final subsection, 
we demonstrate existence of a constant of 
motion of a charged particle associated with a conformal Killing tensor. 
We construct a four-dimensional spacetime which admits 
a non-trivial conformal Killing tensor by a conformal transformation 
of the Minkowski spacetime. 
Making use of the conformal invariance of the Maxwell theory 
in four dimensions, we construct a solution of the Maxwell 
field on the spacetime. 
We show that 
there exists a suitable conformal transformation such that 
the charged particle system has a constant of motion associated with the 
conformal Killing tensor, so that 
the conservation equation holds only on the constraint surface 
in the phase space.

%%%%%%%%%%%%%%%%%%%%%%%%%%%%%%%%%%%%%%%%%%%%%
\subsection{Wald solutions on Kerr geometry}
\label{sec:Wald}
%%%%%%%%%%%%%%%%%%%%%%%%%%%%%%%%%%%%%%%%%%%%%

Let us consider an electro-magnetic field on the Kerr geometry.
If a Ricci flat metric admits a Killing vector, 
the Killing vector solves the vacuum Maxwell equation 
as a test gauge 4-potential in the Lorentz gauge. 
This is called the Wald solution \cite{Wald:1974np}.

The Kerr metric is given by 
\begin{eqnarray}
        ds^2&=&-\left(\frac{\Delta -a^2\sin ^2\theta}{\Sigma}\right)dt^2
                -\frac{2a\sin^2\theta (r^2+a^2-\Delta)}{\Sigma}dt
        d\phi \nonumber\\
        && \quad 
        +\Bigl[ \frac{(r^2+a^2)^2-\Delta a^2 \sin^2 \theta }{\Sigma}\Bigr]
                \sin ^2 \theta d\phi ^2 
        +\frac{\Sigma}{\Delta}dr ^2+ \Sigma d \theta ^2, 
\label{eq:Kerr}
\\
        \Sigma &=& r^2+a^2 \cos^2\theta, 
\label{eq:Sigma}
\\
        \Delta &=& r^2 +a^2 -2M r. 
\label{eq:Delta}
\end{eqnarray}
The spacetime admits two commuting Killing vectors 
$\xi= \partial_t$ and $\psi = \partial _\phi$, 
where $a$ and $M$ are the rotation and mass parameters, respectively. 
On the spacetime, 
\begin{eqnarray}
        A^{\mu} = c_1 \xi^{\mu} + c_2 \psi^{\mu}
        \label{eq-wald}
\label{eq:Waldsolution}
\end{eqnarray}
is a solution of the vacuum Maxwell equations, 
where $c_1$ and $c_2$ are arbitrary constants. 
The solution 
\eqref{eq-wald}
has no magnetic charge. 
In the case $c_1 = 2 a c_2$,  it has no electric charge either. 
Furthermore, the Kerr metric also admits the Killing tensor 
\begin{eqnarray}
        K^{\mu \nu} = 2 \Sigma \,l^{(\mu} n ^{\nu)} + r^2 g^{\mu \nu},
\label{eq:4DKerrKillingtensor}
\end{eqnarray}
which commutes with both of $\xi$ and $\psi$, where 
\begin{eqnarray}
        l^\mu =\frac{r^2+a^2}{\Delta} \xi^\mu
        +\frac{a}{\Delta} \psi^\mu
        +(\partial_{r}){}^{\mu}, 
\qquad
        n^{\mu}=\frac{r^2+a^2}{2\Sigma}\xi^{\mu}
                + \frac{a}{2 \Sigma}\psi^{\mu}
                -\frac{\Delta}{2 \Sigma}(\partial_r)^{\mu}.
\end{eqnarray}

Let us solve the {\eqname} \eqref{eq:N=2EMHEtop}. 
The Killing tensor \eqref{eq:4DKerrKillingtensor}, 
solves the highest-order equation of the hierarchy
\eqref{eq:N=2EMHEtop}. 
We thus set 
\begin{eqnarray}
        \stackrel{(2)}K{}^{\mu \nu}=K^{\mu \nu }. 
\end{eqnarray}
Since $\stackrel{(2)}K$ commutes with $A$, which is a 
linear combination of $\xi$ and $\psi$, 
the second equation \eqref{eq:N=2EMHEmiddle} reduces to the Killing vector equation. 
Thus we have 
\begin{eqnarray}
        \K{1}{}^{\mu}
        =\alpha \xi^{\mu} + \beta \psi^{\mu}, 
\end{eqnarray}
where $\alpha$ and $\beta$ are arbitrary constants. 
Then the last equation \eqref{eq:N=2EMHEbottom} can be written as 
\begin{eqnarray}
        \stackrel{(0)}K_{,\mu}= q^2K_{\mu}^{~\nu} A^2{}_{,\nu}. 
\label{eq:bottomequation}
\end{eqnarray}
By inspecting the integrability condition of the partial differential equation \eqref{eq:bottomequation}, 
we find that this equation is integrable only when $c_2=0$, {\it
  i.e.}, $A=\xi$. 
In the case, the solution is given by 
\begin{eqnarray}
        \K{0} = q^2K_{\mu \nu}\xi^{\mu} \xi ^{\nu}. 
\end{eqnarray}
Thus we have found 
a constant of motion of a charged particle 
associated with a rank-2 Killing tensor, 
$F= (\alpha \xi ^{\mu} + \beta \psi^{\mu}) p_{\mu}
        + K_{\mu \nu}(p_{\mu} p_{\nu}+q^2\xi^{\mu} \xi ^{\nu}). 
$

We conclude that, in the case $A=\xi$,  the 
system has independent Poisson-commuting constants of motion, 
\begin{eqnarray}
        \xi ^{\mu} p_{\mu}, \quad \psi^{\mu} p_{\mu}, \quad \mbox{and}\quad 
        K_{\mu \nu}u^\mu u^\nu, 
        \label{eq-wald-com}
\end{eqnarray}
where $u^\mu:=\frac1m(p^{\mu}-q A^{\mu})$ is the four velocity of the
particle. 
In \eqref{eq-wald-com}, we used the 
fact that $K_{\mu}u^{\nu}u^{\mu}$ is a linear 
combination of $\xi^\mu p_\mu$ and $\psi^\mu p_\mu$, and 
$K^{\mu \nu}(p_{\mu} p_{\nu} +q^2\xi_{\mu} \xi_{\nu})$.
We remark that 
no constant of motion associated with the rank-2 Killing tensor 
exists in the electrically neutral case $c_1 = 2 a c_2$.

%%%%%%%%%%%%%%%%%%%%%%%%%%%%%%%%%%%%
\subsection{Kerr-Newman black holes}
%%%%%%%%%%%%%%%%%%%%%%%%%%%%%%%%%%%%

The Kerr-Newman spacetime is the 
exact solution of electrically charged rotating black hole 
in the Einstein-Maxwell system. 
The spacetime metric is given by \eqref{eq:Kerr} and \eqref{eq:Sigma} with 
\begin{eqnarray}
        \Delta = r^2 +a^2 +e^2 -2M r, 
\label{eq:KNdelta}
\end{eqnarray}
instead of \eqref{eq:Delta}, and electro-magnetic 4-potential is given by
\begin{eqnarray}
        A &=&-\frac{e r}{\Sigma}( dt -a\sin ^2 \theta d \phi), 
\label{eq:gaugepotential}
\end{eqnarray}
where $e$ is electric charge of the black hole. 
As the Kerr metric, the Kerr-Newman metric admits two Killing vectors 
$\xi=\partial_t$ and $\psi=\partial_r$ and 
the Killing tensor $K^{\mu \nu}$ in 
\eqref{eq:4DKerrKillingtensor}, 
where $\Delta$ in $l$ and $n$ is now given by \eqref{eq:KNdelta}.

In a manner similar to that in the previous section, 
we can find the solution to the 
set of equations \eqref{eq:N=2EMHEtop}, \eqref{eq:N=2EMHEmiddle}, 
and \eqref{eq:N=2EMHEbottom}, 
\begin{align}
&\K{2}_{\mu \nu }=K_{\mu \nu }, \\
&\K{1}_{\mu}=-2q K^{\nu}{}_{\mu}A_{\nu} + \alpha \xi_{\mu} + \beta \psi_{\mu}, \\
&\K{0}=q^2K^{\mu \nu}A_{\mu} A_{\nu},
\end{align}
where $\alpha$ and $\beta$ are arbitrary constants. 
Then the constant of motion associated with the Killing tensor is given by
\begin{align}
        F&=K^{\mu \nu }p_{\mu}p_{\nu}
                +(-2qK^{\mu \nu}A_{\nu} + \alpha \xi^{\mu} + \beta \psi^{\mu})p_{\mu}
                +q^2K^{\mu \nu }A_{\mu}A_{\nu}\cr
        &= K_{\mu \nu }u^{\mu}u^{\nu} + \alpha \xi^{\mu}p_{\mu} + \beta\psi^{\mu}p_{\mu}.
\end{align}
As the Wald solution $A=\xi$ on the Kerr metric discussed in the previous
section,  
the Kerr-Newman metric admits 
the independent constants of motion, 
$\xi ^{\mu} p_{\mu}, \psi^{\mu} p_{\mu}$, and $K_{\mu \nu}u^{\mu}u^{\nu}$, 
for a charged particle. 
These two examples have common properties, {\it i.e.}, the both are 
rotating black holes having an electric monopole and a 
magnetic field falling off toward infinity. 

The constant of motion for a charged particle 
associated with the Killing tensor is referred to 
as Carter's constant, which was first obtained by 
the Hamilton-Jacobi method in Ref.~\cite{Carter:1968rr}.

%%%%%%%%%%%%%%%%%%%%%%%%%%%%%%%%%%%%%%%%%%%%%%%%%
\subsection{Five-dimensional Charged Black Holes}
\label{sec:sec4-3}
%%%%%%%%%%%%%%%%%%%%%%%%%%%%%%%%%%%%%%%%%%%%%%%%%

We demonstrate that our formalism is applicable to 
a charged particle moving around a five-dimensional charged black hole. 
Here, we consider the five-dimensional charged rotating black hole 
with the following 
metric and the electro-magnetic 5-potential \cite{Chong:2005hr}, 
\begin{eqnarray}
        ds^2 &=& -\frac{ (\rho^2 d t + 2e \nu)\, d t}{\rho^2} 
                + \frac{2e\, \nu\omega}{\rho^2}
        + \frac{f}{\rho^4}\Big( d t -\omega\Big)^2 
        + \frac{\rho^2 d r^2}{\Delta_r} 
        +\rho^2 d\theta^2 
\nonumber\\
&& + (r^2+a^2 )\sin^2\theta d\phi^2 +
(r^2+b^2 )\cos^2\theta d\psi^2,
\\
        A &=& \frac{\sqrt3 e}{2 \rho^2}\,
\Big( d t
- a\sin^2\theta d\phi - b\cos^2\theta d\psi\Big),
\end{eqnarray}
where 
\begin{eqnarray}
        &&S = a^2 \cos ^2 \theta + b^2 \sin ^2 \theta, \quad
        \rho^2 = r^2 + S, \quad
        f = 2 M \rho^2 - e^2,
\\
        &&\nu = b\sin^2\theta d\phi + a\cos^2\theta d\psi, \quad 
        \omega = a\sin^2\theta d\phi + b\cos^2\theta d\psi,
\\
        &&\Delta_r = \frac{(r^2+a^2)(r^2+b^2) + e^2 +2ab e}{r^2} - 2M. 
\end{eqnarray}
The black hole is characterized by mass parameter $M$, charge parameter $e$, 
and two spin parameters $a$ and $b$. 
The metric is an exact solution 
in the five-dimensional Einstein-Maxwell-Chern-Simons theory. 
The metric admits three Killing vectors 
$\partial_t$, $\partial_\phi$, and $\partial_\psi$, and an irreducible 
Killing tensor \cite{Davis:2005ys} 
\begin{eqnarray}
K ^{\mu\nu} \ =\ -S g^{\mu\nu} -S (\partial_t)^{\mu} (\partial_t)^{\nu}
+ \frac{1}{\sin^2\theta} (\partial_{\phi})^{\mu} (\partial_{\phi})^{\nu}+
\frac{1 }{\cos^2\theta} (\partial_{\psi})^{\mu} (\partial_{\psi})^{\nu} +
(\partial_{\theta})^{\mu} (\partial_{\theta})^{\nu}. 
\label{eq:5DBHKT}
\end{eqnarray}
These Killing vectors and tensor commute with each other. 
Hence, we can discuss the existence of constants of motion of 
a charged particle associated with the three Killing vectors and 
the Killing tensor.

Let us consider the {\eqname} \eqref{eq:N=2EMHEtop}, 
\eqref{eq:N=2EMHEmiddle}, and \eqref{eq:N=2EMHEbottom}, 
with the rank-2 Killing tensor 
$\K{2}{}^{\mu \nu} = K^{\mu \nu}$ which solves \eqref{eq:N=2EMHEtop}. 
We try to find the solution of the second equations \eqref{eq:N=2EMHEmiddle} 
of the form 
\begin{eqnarray}
        \stackrel{(1)}K{}^{\mu}{}^{; \nu} 
                + \stackrel{(1)}K{}^{\nu}{}^{; \mu}
        = B^{\mu \nu}, 
\label{eq:hierarchy(5DBH)}
\end{eqnarray}
where 
$B^{\mu \nu}
        =2 q\left( A^{\lambda}K{}^{\mu \nu}{}_{,\lambda}
                -2K{}^{\lambda (\mu}A^{\nu)}{}_{,\lambda}\right).
$
Since $t$, $\phi$, and $\psi$ are the Killing coordinates, 
the equation \eqref{eq:hierarchy(5DBH)} 
has a simple form, 
which is shown in appendix \ref{sec:app2}.

To find solutions of \eqref{eq:N=2EMHEmiddle}, 
we assume that $\K{1}$ depends only on $r$ and $\theta$ 
as $\K{2}$ does. It turns out that 
\begin{eqnarray}
\K{1}{}^{r} = 0, 
\quad \K{1}{}^{\theta}= 0 
\end{eqnarray}
from the explicit form of \eqref{eq:hierarchy(5DBH)}, 
and the other components of $\K{1}$ satisfy the following equations, 
\begin{eqnarray}
        g^{rr}\K{1}{}^{t}{}_{,r} &=& B^{tr}, \quad
        g^{\theta\theta}\K{1}{}^{t}{}_{,\theta} = B^{t\theta}, \quad
        g^{rr}\K{1}{}^{\phi}{}_{,r} = B^{r\phi}, 
\nonumber\\
        g^{\theta\theta}\K{1}{}^{\phi}{}_{,\theta} &=& B^{\theta\phi},\quad
        g^{rr}\K{1}{}^{\psi}{}_{,r} = B^{r\psi}, \quad
        g^{\theta\theta}\K{1}{}^{\psi}{}_{,\theta} = B^{\theta\psi}. 
\end{eqnarray}
After some calculations,
we can find an explicit solution 
\begin{eqnarray}
  \stackrel{(1)}K{}^{\mu} 
  = 2qS A^{\mu} 
  +\alpha (\partial_t)^\mu +\beta (\partial_\phi)^\mu 
  +\gamma (\partial_\psi)^\mu, 
\end{eqnarray}
where $\alpha, \beta$ and $\gamma$ are arbitrary constants.

Let us solve \eqref{eq:N=2EMHEbottom}. 
It can be rewritten as 
\begin{eqnarray}
&&\stackrel{(0)}K{}_{,t}=\stackrel{(0)}K{}_{,\phi}=\stackrel{(0)}K{}_{,\psi} =0, 
\\
&&g^{r r}\,\stackrel{(0)}K{}_{,r} 
        =-\frac{3 q^2e^2 S}{2 r^3 \Delta _r \rho^6} 
        \left[ 
        ( r^4 - (ab +e)^2 ) S + r^4 \left( (a^2+b^2) + 2 (r^2 - M) \right) 
        \right], 
\\
        &&g^{\theta \theta}\,\stackrel{(0)}K{}_{,\theta}
        =-\frac{3q^2 e^2 r^2}{4 \Delta _r \rho ^{6}}(a+b)(a-b) \sin 2\theta.
\end{eqnarray}
We can easily integrate these equations to get 
\begin{eqnarray}
        \stackrel{(0)}K = - q^2S A^{\mu} A_{\mu}. 
\end{eqnarray}

As a result, we obtain a constant of motion associated with the Killing 
tensor, 
\begin{eqnarray}
        F = K^{\mu \nu} p_{\mu} p_{\nu} + 2 qS A^{\mu} p_{\mu} - q^2S A^{\mu}A_{\nu} 
\end{eqnarray}
for a charged particle moving in the five-dimensional charged rotating black
holes in addition to the momentum components $p_t, p_\phi$ and $p_\psi$. 
These four constants of motions Poisson-commute mutually.

%%%%%%%%%%%%%%%%%%%%%%%%%%%%%%%%%%%%%%%%%%%%%%%%%%%%%%%%%%%%%%%%%%%%%%%%
\subsection{Constant of Motion associated with Conformal Killing Tensor}
%%%%%%%%%%%%%%%%%%%%%%%%%%%%%%%%%%%%%%%%%%%%%%%%%%%%%%%%%%%%%%%%%%%%%%%%

In this subsection, we demonstrate that the conservation equation can hold only on the constraint surface. 
In such a case, the constant of motion is
associated 
with a conformal Killing tensor. 
Unfortunately, we have not found an example for this case 
as a solution of the Einstein equation, but we present some example 
of curved spacetime. 

Let $({\cal M}^4, \bar g)$ be the Minkowski spacetime. 
Consider the metric $\bar g$ spanned by the polar coordinates as
\begin{equation}
        {d\bar s}^2 = \bar g_{\mu\nu}dx^\mu dx^\nu 
                =-dt^2 + dr^2 + r^2(d\theta^2 + \sin^2\theta d\phi^2). 
\end{equation}
The Minkowski spacetime admits a Killing tensor 
\begin{eqnarray}
        K^{\mu \nu } = (\partial{}_{\theta})^{\mu} 
        (\partial_{\theta})^{\nu} 
        + \frac{1}{\sin^2 \theta} (\partial{}_{\phi})^{\mu}
        (\partial{}_{\phi})^{\nu}. 
\label{eq:KT_in_Minkowski}
\end{eqnarray}
As a solution of the test electro-magnetic field on the flat background, 
we take $\bar A = \partial_{\phi}$. 
This is the special case of the Wald solution discussed in sec.~\ref{sec:Wald}, 
with $M=0$ and $a=0$. 
As was mentioned there, there is no constant of motion associated 
with the Killing tensor $K^{\mu\nu}$ for a charged particle. 

Let $({\cal M}^4,g)$ be 
a spacetime 
with $g_{\mu \nu} = e^{\Phi} \bar g_{\mu \nu}$ 
where $\Phi$ is a function on ${\cal M}^4$. 
The spacetime $({\cal M}^4,g)$ is conformally flat. 
Since the Maxwell theory in a four-dimensional spacetime 
has conformal invariance, 
the gauge 4-potential 
$A_\mu  := \bar A_\mu $ 
solves 
the Maxwell equations on $({\cal M}^4,g)$. 
The tensor $K^{\mu\nu}$ given by \eqref{eq:KT_in_Minkowski}, satisfying 
\begin{eqnarray}
        - \sbr{K}{g} = \stp{2 \kphi}{g},
\end{eqnarray}
is a conformal Killing tensor on $({\cal M}^4,g)$, 
where $K\partial$ denotes the derivative operator $K^{\mu\nu}\partial_\nu$. 
We show that there can exist a constant of motion associated with 
the conformal Killing tensor for a class of the conformal factor.

We shall try to find a solution of the {\eqname} which starts with a rank-2 
conformal Killing tensor which is not a Killing tensor. 
Namely, we shall solve \eqref{eq:EMHEtop}-\eqref{eq:EMHE} with 
\eqref{eq:generalconsistencycondition}, {\it i.e.}, 
\begin{align}
&       - \sbr{\K{2}}{g} + \stp{\T{\lambda}{1}}{g} =0,
\label{eq:nonzerolambdaN=2EMHEtop}
\\&
- \sbr{\K{1}}{g} + 2q \sbr{\K{2}}{A} 
+\T{\lambda}{0} g -2 q \stp{\T{\lambda}{1}}{A} =0,
\label{eq:nonzerolambdaN=2EMHEmiddle}
\\&
- \sbr{\K{0}}{g} + 2 q \sbr{\K{1}}{A} 
-q^2 \sbr{\K{2}}{A^2} -2 q \T{\lambda}{0} A 
+\T{\lambda}{1}(q^2 A^2 + m^2) =0,
\label{eq:nonzerolambdaN=2EMHEbottom}
\end{align}
with 
\begin{align}
        m^{2} ( \T{\lambda}{0} + q \T{\lambda}{1} \cdot \T{A}{1}) =0, 
\label{eq:N=2consistencycondition}
\end{align}
where $\T{\lambda}{1}$ is not identically zero 
and the gauge potential $A^\mu$ is given by 
$
A^{\mu} = e^{- \Phi} (\partial_{\phi})^\mu. 
$

The tensor $\K{2}{}^{\mu\nu}=K^{\mu\nu}$ given by \eqref{eq:KT_in_Minkowski} solves 
\eqref{eq:nonzerolambdaN=2EMHEtop} 
with 
\begin{eqnarray}
        \T{\lambda}{1} = - 2 \kphi. 
\end{eqnarray}
From the algebraic condition \eqref{eq:N=2consistencycondition}, we have 
\begin{eqnarray}
\T{\lambda}{0} 
= - q \ip{A}{\T{\lambda}{1}} 
= 2q e^{- \Phi} r^2 \partial _{\phi}\Phi. 
\end{eqnarray}
We assume that the function $\Phi$ is in the form $\Phi = \Phi(t,r,\theta)$, 
so that we have $\T{\lambda}{0}=0$ and $\sbr{K}{A}=0 $. 
Then the second equation  
\eqref{eq:nonzerolambdaN=2EMHEmiddle} 
reduces to the Killing vector equation 
\begin{equation}
        \sbr{\K{1}}{g} = 0. 
\end{equation}
For simplicity, we choose the trivial solution $\K{1} = 0$. 
The remaining 
equation \eqref{eq:nonzerolambdaN=2EMHEbottom} 
is given by 
\begin{eqnarray}
        \sbr{\K{0}}{g} =
        - 2 q^2 e^{-\Phi} K\partial \bar g_{\phi\phi} -2m^2 \kphi, 
\end{eqnarray}
or explicitly, 
\begin{eqnarray}
        &&\partial_t \K{0} = \partial_r \K{0}= \partial_\phi \K{0}= 0, 
\label{eq:K00}\\
        &&\partial_\theta \K{0}
                =\bar g_{\theta\theta}(q^2 \partial_\theta \bar g_{\phi\phi} 
                +m^2 e^{\Phi} \partial_\theta \Phi). 
\label{eq:K0}
\end{eqnarray}
There exists a constant of motion if the function $\Phi$ is chosen 
such that the partial differential equations \eqref{eq:K00} and 
\eqref{eq:K0} 
for $\K{0}$ are integrable. 

As the simplest case, 
we consider that the right-hand side of \eqref{eq:K0} vanishes, namely, 
\begin{align} 
        m^2 \partial _\theta e^{\Phi}
                +2q^2 r^2 \sin \theta \cos \theta =0.
\label{eq:eqPhi}
\end{align}
In this case, equations \eqref{eq:K00} and \eqref{eq:K0} 
admit a trivial solution $\K{0} = 0$. 
We can integrate \eqref{eq:eqPhi} easily to obtain 
\begin{eqnarray}
        e^{\Phi} =\frac{q^2}{m^2} r^2 \cos ^2 \theta + f(t,r), 
\label{eq:cfactor}
\end{eqnarray}
where $f(t,r)$ is an arbitrary positive function.

Therefore, if we choose \eqref{eq:cfactor} as the conformal factor, 
then the quadratic quantity $F = K \cdot p^2$ 
 is a constant of motion 
of a charged particle associated with the conformal Killing tensor. 
The quantity is conserved only on the constraint surface in the phase space.

%%%%%%%%%%%%%%%%%%%%%%%%%%%%%%%%%%%%%%%%%%%%%%
\section{Summary}
\label{sec:sec5}
%%%%%%%%%%%%%%%%%%%%%%%%%%%%%%%%%%%%%%%%%%%%%%

In this paper, we have discussed constants of motion for a test particle in a curved spacetime. 
For the particle which is subjected to an external field 
we have obtained the condition for existence of the constant of motion 
in the form of coupled equations with a hierarchical structure. 
There, we have taken the Hamiltonian constraint 
condition for the particle, which arises from the reparametrization 
invariance of particle's world line, into consideration. 
The equation at the top of the hierarchy is the conformal Killing tensor equation. 
Then, the existence of constant of motion requires that the metric admits a 
conformal Killing tensor. If the {\eqname} has a non-trivial solution, a constant of motion associated with the conformal Killing tensor exists.

As applications of the formalism, we have considered systems of a charged particle 
in Maxwell's fields on black holes. 
In the case of a charged particle in the Kerr-Newman black holes, we have 
rediscovered a constant of motion quadratic in the canonical momenta, 
which has been found via the 
Hamilton-Jacobi method \cite{Carter:1968rr}.
In the case of a charged particle in the electro-magnetic field without 
electric charge constructed by Wald's method on the Kerr black holes, 
we have shown that the {\eqname} is not integrable. 
The non-existence of constant of motion does 
not depend on the 
choice of coordinate, in contrast to the fact that the discovery of 
constant of motion was due to suitable coordinates in the Hamilton-Jacobi method. 
We have found a new constant of motion as a solution of the hierarchical 
equations for a charged particle around a 
five-dimensional charged rotating black hole, which is a solution for 
the Einstein-Maxwell-Chern-Simons theory. 
Since these metrics admit Killing tensor of rank 2, the constants of motion in these examples are associated with the Killing tensors. 

As the final example in this paper, we have considered Maxwell's field 
on an artificial conformal flat spacetime. 
For a charged particle in this fields, we constructed a constant of motion 
which is associated with rank-2 conformal Killing tensor, 
{\it i.e.}, conservation equation holds only on the Hamiltonian constraint surface in the phase space. 
It would be interesting problem to find constants of motion for a particle 
moving in a solution of the Einstein-Maxwell system. 
The extension to a wide classes of interactions is an important future work.

%%%%%%%%%%%%%%%%%%%%%%%%%%%%%%%%%%
\section*{Acknowledgements}
{
This work is supported in part by Keio Gijuku Academic Development Funds (T.K.) 
and the Grant-in-Aid for Scientific Research No.19540305 (H.I.).
}

%%%%%%%%%%%%%%%%%%%%%%%%%%%%%%%%%%%%%%%%%%%%
\appendix
\section{{\eqname} for a free particle}
\label{sec:app1}

Let us consider the system of a free particle in the framework of 
the {\eqname}.
The explicit form of the Hamiltonian is given by \eqref{eq:Hgeo} 
and the constraint equation is given by \eqref{eq:Hamiltonian_Constraint}. 
Then the {\eqname} \eqref{eq:general_hierarchy}
reads  
\begin{equation}
\begin{aligned}
        &- \sbr{\K{k-1}}{g} +\stp{\T{\lambda}{k-2}}{g} 
                +m^2 \T{\lambda}{k} = 0, \quad k\geq2, 
\cr
        &- \sbr{\K{0}}{g} + m^2 \T{\lambda}{1} = 0, 
\cr
        &\quad m^2 \T{\lambda}{0} = 0.
\label{eq:geodHE}
\end{aligned}
\end{equation}

If the particle is massless, {\it i.e.}\/ $m = 0$, 
then the constraint 
\eqref{eq:Hamiltonian_Constraint} 
becomes
\begin{eqnarray}
        {\cal H} = g^{\mu \nu}p_{\mu} p_{\nu} \approx 0, 
\label{eq:massless}
\end{eqnarray}
and the {\eqname} becomes 
\begin{equation}
\begin{aligned}
        &- \sbr{\K{k-1}}{g} +\stp{\T{\lambda}{k-2}}{g} = 0, 
        \quad k\geq2, 
\cr
        &- \sbr{\K{0}}{g} = 0.
\label{eq:nullHE}
\end{aligned}
\end{equation}
All the equations for $\K{k}~ (k\geq 1)$ become decoupled 
conformal Killing equations. 
Therefore a nontrivial conformal Killing tensor with non-vanishing 
$\T{\lambda}{l}$ gives 
a constant of motion conserved only on the constraint 
surface \eqref{eq:massless}.

If the particle is massive, {\it i.e.}\/ $m \neq 0$, 
we consider symmetric tensors $\stackrel{(k)}{\widetilde{K}}$ 
which satisfy the linear differential equations
\begin{align}
        -\sbr{\T{{\widetilde K}}{k}}{g} + m^2 \T{\lambda}{k+1} =0. 
\label{eq:element}
\end{align}
The solutions of the linear differential equations \eqref{eq:element} have the form
\begin{align}
\T{{\widetilde K}}{k}
= \T{{\widetilde K}_{\rm H}}{k}
+ \T{{\widetilde K}_{\rm I}}{k},
\end{align}
where the homogeneous part $\T{{\widetilde K}_{\rm H}}{k}$ is a solution of 
the Killing equation
\begin{align}
        \sbr{\T{{\widetilde K}_{\rm H}}{k}}{g}=0, 
\end{align}
and $\T{{\widetilde K}_{\rm I}}{k}$ is an inhomogeneous solution 
satisfying the original equations \eqref{eq:element}, 
\begin{eqnarray}
        -\sbr{\T{{\widetilde K}_{\rm I}}{k}}{g} + m^2 \T{\lambda}{k+1} =0. 
\end{eqnarray}
Using $\tilde K_{\rm H}$ and $\tilde K_{\rm I}$, 
we can construct the solution of \eqref{eq:geodHE} as
\begin{align}
        \T{K}{k}
                = \T{{\widetilde K}_{\rm H}}{k}+\T{{\widetilde K}_{\rm I}}{k}
                + \frac{1}{m^2}\stp{\T{{\widetilde K}_{\rm I}}{k-2}}{g}.
\end{align}
For this solution, the conserved quantity has the form 
\begin{eqnarray}
        F
        &=& \sum_{k} \ip{\K{k}}{p^{k}} 
\cr
        &=&\sum_{k} 
        \Big(
        \ip{\T{{\widetilde K}_{\rm H}}{k}}{p^k}
        + \ip{\T{{\widetilde K}_{\rm I}}{k}}{p^k} 
        \Big)
        + \sum_{k} 
        \ip{\Big(\frac{1}{m^2}\stp{\T{{\widetilde K}_{\rm I}}{k-2}}{g}\Big)}{p^{k}} 
\cr
        &=&\sum_{k}\ip{\T{{\widetilde K}_{\rm H}}{k}}{p^k}
+\frac{1}{m^2}\sum_{k}
\ip{\T{{\widetilde K}_{\rm I}}{k}}{p^k} 
                        (\ip{g}{p^{2}}+m^2) \cr
        &\approx& \sum_{k}^{}
\ip{\T{{\widetilde K}_{\rm H}}{k}}{p^k}. 
\end{eqnarray} 
Therefore the homogeneous solutions, that is, solutions for the Killing equations, contribute to the conserved quantity. 
The constant of motion for a massive free particle is requires the existence of 
the Killing tensor.

%%%%%%%%%%%%%%%%%%%%%%%%%%%%%%%%%%%%%%%%%%%%%%%%%%%%%%%%%%%%%
\section{Equations in the case of five-dimensional black holes}
\label{sec:app2}

As a supplement to section~\ref{sec:sec4-3}, 
we give the explicit form of equations \eqref{eq:hierarchy(5DBH)} in terms of
components: 
\begin{align}
(t,r):&\qquad
g^{rr}\K{1}{}^{t}{}_{,r}
+ g^{tt}\K{1}{}^{r}{}_{,t}
+ g^{\phi t}\K{1}{}^{r}{}_{,\phi}
+ g^{\psi t}\K{1}{}^{r}{}_{,\psi}
= B^{tr},&
\\
(t,\theta):&\qquad
g^{\theta\theta}\K{1}{}^{t}{}_{,\theta}
+ g^{tt}\K{1}{}^{\theta}{}_{,t}
+ g^{\phi t}\K{1}{}^{\theta}{}_{,\phi}
+ g^{\psi t}\K{1}{}^{\theta}{}_{,\psi}
= B^{t\theta},&
\\
(r,\phi):&\qquad
g^{t\phi}\K{1}{}^{r}{}_{,t}
+ g^{\phi\phi}\K{1}{}^{r}{}_{,\phi}
+ g^{\psi\phi}\K{1}{}^{r}{}_{,\psi}
+ g^{rr}\K{1}{}^{\phi}{}_{,r}
= B^{r\phi},&
\\
(r,\psi):&\qquad
g^{t\psi}\K{1}{}^{r}{}_{,t}
+ g^{\phi\psi}\K{1}{}^{r}{}_{,\phi}
+ g^{\psi\psi}\K{1}{}^{r}{}_{,\psi}
+ g^{rr}\K{1}{}^{\psi}{}_{,r}
= B^{r\psi},&
\\
(\theta,\phi):&\qquad
g^{t\phi}\K{1}{}^{\theta}{}_{,t}
+ g^{\phi\phi}\K{1}{}^{\theta}{}_{,\phi}
+ g^{\psi\phi}\K{1}{}^{\theta}{}_{,\psi}
+ g^{\theta\theta}\K{1}{}^{\phi}{}_{,\theta}
= B^{\theta\phi},&
\\
(\theta,\psi):&\qquad
g^{t\psi}\K{1}{}^{\theta}{}_{,t}
+ g^{\phi\psi}\K{1}{}^{\theta}{}_{,\phi}
+ g^{\psi\psi}\K{1}{}^{\theta}{}_{,\psi}
+ g^{\theta\theta}\K{1}{}^{\psi}{}_{,\theta}
= B^{\theta\psi},&
\end{align}
where components of $B^{\mu \nu}$ are given explicitly by 
\begin{align}
B^{t \theta}
        &=
\frac{\sqrt{3} qe }{\Delta _r \rho^6}\left[(r^2+a^2)(r^2+b^2)+abe\right]
(a+b)(a-b)\sin 2 \theta,
\\
B^{\theta \phi}
        &=
\frac{\sqrt3 q e}{\Delta _{r} \rho^{6}} 
\left[b(ab+e)+ar^2 \right](a+b)(a-b)\sin 2 \theta,
\\
B^{\theta \psi}
        &=
\frac{\sqrt3 qe}{\Delta _{r} \rho^{6}}
\left[a(ab+e)+br^2 \right](a+b)(a-b)\sin 2 \theta,
\\
B^{tr} 
        &=
\frac{2 \sqrt{3} qe S}{r^3 \Delta _r \rho^6} 
\biggl[ 
\left[(r^2+a^2)(r^2+b^2)+abe\right]
\left[r^2 \Delta _r + \rho^2 (a^2 + b^2 
+ 2(r^2- M))\right]
\cr
        & \hspace{3cm} 
-r^2 \rho^2 \Delta _r (a^2 + b^2 + 2r^2)
\biggr], 
\\
B^{r\phi} 
        &=
\frac{2 \sqrt{3} qe S}{r^3 \Delta _r \rho^6} 
\left[ 
\left( b(e+ab)+ar^2 \right)
\left[ r^2 \Delta _r + \rho^2 (a^2 +b^2 +2(r^2-M)) \right] 
- a r ^2 \Delta _r \rho^2
\right],
\\
B^{r\psi} 
        &=
\frac{2 \sqrt{3} qe S}{r^3 \Delta _r \rho^6} 
\biggl[ 
\left( a(e+ab)+br^2 \right)
\left[ r^2 \Delta _r + \rho^2 (a^2 +b^2 +2(r^2-M)) \right] 
- b r ^2 \Delta _r \rho^2
\biggr].
\end{align}
The other components of the equation are trivial.

%%%%%%%%%%%%%%%%%%%%%%%%%%%%%%%%%%%%%%%%
\section*{references}


\begin{thebibliography}{99}
%\cite{Carter:1968rr}
\bibitem{Carter:1968rr}
B.~Carter,
%``Global structure of the Kerr family of gravitational fields,''
Phys.\ Rev.\ {\bf 174}, 1559 (1968).
%%CITATION = PHRVA,174,1559;%%


%\cite{Walker:1970un}
\bibitem{Walker:1970un}
M.~Walker and R.~Penrose,
%``On quadratic first integrals of the geodesic equations for type [22]
%spacetimes,''
Commun.\ Math.\ Phys.\ {\bf 18}, 265 (1970).
%%CITATION = CMPHA,18,265;%%


\bibitem{Frolov:2007nt}
V.~P.~Frolov and D.~Kubiznak,
%``'Hidden' symmetries of higher dimensional rotating black holes,''
Phys.\ Rev.\ Lett.\ {\bf 98}, 011101 (2007)
[arXiv:gr-qc/0605058].
%%CITATION = PRLTA,98,011101;%%


\bibitem{Kubiznak:2006kt}
D.~Kubiznak and V.~P.~Frolov,
%``Hidden Symmetry of Higher Dimensional Kerr-NUT-AdS Spacetimes,''
Class.\ Quant.\ Grav.\ {\bf 24}, F1 (2007)
[arXiv:gr-qc/0610144].
%%CITATION = CQGRD,24,F1;%%

\bibitem{Page:2006ka}
D.~N.~Page, D.~Kubiznak, M.~Vasudevan and P.~Krtous,
%``Complete Integrability of Geodesic Motion in General Kerr-NUT-AdS
%Spacetimes,''
Phys.\ Rev.\ Lett.\ {\bf 98}, 061102 (2007)
[arXiv:hep-th/0611083].
%%CITATION = PRLTA,98,061102;%%

\bibitem{Frolov:2006pe}
V.~P.~Frolov, P.~Krtous and D.~Kubiznak,
%``Separability of Hamilton-Jacobi and Klein-Gordon equations in general
%Kerr-NUT-AdS spacetimes,''
JHEP {\bf 0702}, 005 (2007)
[arXiv:hep-th/0611245].
%%CITATION = JHEPA,0702,005;%%

\bibitem{Krtous:2006qy}
P.~Krtous, D.~Kubiznak, D.~N.~Page and V.~P.~Frolov,
%``Killing-Yano tensors, rank-2 Killing tensors, and conserved quantities in
%higher dimensions,''
JHEP {\bf 0702}, 004 (2007)
[arXiv:hep-th/0612029].
%%CITATION = JHEPA,0702,004;%%

\bibitem{Krtous:2007xf}
P.~Krtous, D.~Kubiznak, D.~N.~Page and M.~Vasudevan,
%``Constants of Geodesic Motion in Higher-Dimensional Black-Hole Spacetimes,''
Phys.\ Rev.\ D {\bf 76}, 084034 (2007)
[arXiv:0707.0001 [hep-th]].
%%CITATION = PHRVA,D76,084034;%%


%\cite{Hughston:1972qf}
\bibitem{Hughston:1972qf}
L.~P.~Hughston, R.~Penrose, P.~Sommers and M.~Walker,
%``On A Quadratic First Integral For The Charged Particle Orbits In The
%Charged Kerr Solution,''
Commun.\ Math.\ Phys.\ {\bf 27}, 303 (1972).
%%CITATION = CMPHA,27,303;%%

%\cite{P.Sommers:1973qf}
\bibitem{P.Sommers:1973qf}
P.~Sommers 
%``On Killing tensors and constants of motion,''
J.\ Math.\ Phys.\ {\bf 14}, 787 (1973).
%%CITATION = %%


%\cite{vanHolten:2006xq}
\bibitem{vanHolten:2006xq}
J.~W.~van Holten,
%``Covariant Hamiltonian dynamics,''
Phys.\ Rev.\ D {\bf 75}, 025027 (2007)
% [arXiv:hep-th/0612216].
%%CITATION = PHRVA,D75,025027;%%

%\cite{Schouten}
\bibitem{Schouten}
J. Schouten: Ricci Calculus, Springerverlag Berlin (1954).

%%%%%%%%%%%%
\bibitem{Hierarchy}
%\bibitem{Ngome:2010gg}
J.~P.~Ngome, P.~A.~Horvathy and J.~W.~van Holten,
%``Dynamical supersymmetry of spin particle-magnetic field interaction,''
arXiv:1003.0137 [hep-th];
%%CITATION = ARXIV:1003.0137;%%

%\bibitem{Visinescu:2009rm}
M.~Visinescu,
%``Higher order first integrals of motion in a gauge covariant Hamiltonian
%framework,''
Mod.\ Phys.\ Lett.\ A {\bf 25}, 341 (2010)
[arXiv:0910.3474 [hep-th]];
%%CITATION = MPLAE,A25,341;%%

%\bibitem{Ngome:2009pa}
J.~P.~Ngome,
%``Curved manifolds with conserved Runge-Lenz vectors,''
J.\ Math.\ Phys.\ {\bf 50}, 122901 (2009)
[arXiv:0908.1204 [math-ph]];
%%CITATION = JMAPA,50,122901;%%

%\bibitem{Horvathy:2009nq}
P.~A.~Horvathy and J.~P.~Ngome,
%``Conserved quantities in a non-abelian monopole field,''
Phys.\ Rev.\ D {\bf 79}, 127701 (2009)
[arXiv:0902.0273 [hep-th]].
%%CITATION = PHRVA,D79,127701;%%
%%%%%%%%%%%%%%%%%%%%%%%%%


%\cite{Wald:1974np}
\bibitem{Wald:1974np}
R.~M.~Wald,
%``Black hole in a uniform magnetic field,''
Phys.\ Rev.\ D {\bf 10}, 1680 (1974).
%%CITATION = PHRVA,D10,1680;%%


%\cite{Chong:2005hr}
\bibitem{Chong:2005hr}
Z.~W.~Chong, M.~Cvetic, H.~Lu and C.~N.~Pope,
%``General non-extremal rotating black holes in minimal five-dimensional
%gauged supergravity,''
Phys.\ Rev.\ Lett.\ {\bf 95}, 161301 (2005)
% [arXiv:hep-th/0506029].
%%CITATION = PRLTA,95,161301;%%


%\cite{Davis:2005ys}
\bibitem{Davis:2005ys}
P.~Davis, H.~K.~Kunduri and J.~Lucietti,
%``Special symmetries of the charged Kerr-AdS black hole of D = 5 minimal
%gauged supergravity,''
Phys.\ Lett.\ B {\bf 628}, 275 (2005)
[arXiv:hep-th/0508169].
%%CITATION = PHLTA,B628,275;%%

\end{thebibliography}
\end{document}